\documentclass[aps,twocolumn,prd]{revtex4}

\usepackage{latexsym}
\usepackage{epsfig}
\usepackage{scrextend}

\def\be{\begin{equation}}
\def\ee{\end{equation}}
\def\bea{\begin{eqnarray}}
\def\eea{\end{eqnarray}}
\def\ba{\begin{array}} 
\def\ea{\end{array}}
\def\bc{\begin{center}}
\def\ec{\end{center}}

\def\ghost#1{}

\def\simge{\mathrel{%
   \rlap{\raise 0.511ex \hbox{$>$}}{\lower 0.511ex \hbox{$\sim$}}}}
\def\simle{\mathrel{
   \rlap{\raise 0.511ex \hbox{$<$}}{\lower 0.511ex \hbox{$\sim$}}}}

\def\dis{\displaystyle}

\def\be{\begin{equation}}
\def\ee{\end{equation}}
\def\bea{\begin{eqnarray}}
\def\eea{\end{eqnarray}}

\def\ba{\begin{array}} 
\def\ea{\end{array}}
\def\bc{\begin{center}}
\def\ec{\end{center}}

\def\ghost#1{}

\def\simge{\mathrel{%
   \rlap{\raise 0.511ex \hbox{$>$}}{\lower 0.511ex \hbox{$\sim$}}}}
\def\simle{\mathrel{
   \rlap{\raise 0.511ex \hbox{$<$}}{\lower 0.511ex \hbox{$\sim$}}}}

\begin{document}

\title{\boldmath
 The  BE-Higgs boson as spin-0 partner of the $Z$ \vspace{3.5mm}\\
 in the Supersymmetric Standard Model\vspace{10mm}\\}

\vspace{-3mm}

\author{Pierre FAYET\vspace{5mm}
\\}

\vspace{0mm}

\address{Laboratoire de Physique Th\'eorique de l'ENS
{\small \ (UMR 8549 CNRS)} \\
 24 rue Lhomond, 75231 Paris Cedex 05, France \vspace{1mm}\\}

\begin{abstract}

\textwidth 18cm

Supersymmetric extensions of the standard model
 lead to gauge/BE-Higgs unification by providing spin-0 bosons as {\it extra states for spin-1 gauge bosons} 
within massive gauge multiplets. They may be described by the spin-0 components of 
{\it \,massive gauge superfields} \ (instead of chiral superfields as usual).

\vspace{.3mm}
In particular, the  125 GeV$/c^2$ boson observed at CERN, considered as a BEH boson associated 
with electroweak breaking and mass generation, may also be interpreted, up to a mixing angle induced by supersymmetry breaking, as
{\it the spin-0 partner of the $Z$} under {\it two} supersymmetry transformations, 
i.e. as a  $Z$ that would be deprived of its spin.
\vspace{-1.5mm}\\

\end{abstract}

\maketitle

\section{Introduction}

\vspace{-1mm}

Supersymmetric extensions of the standard model lead to {\it superpartners}  for all particles,   {squarks} and {sleptons},
 {gluinos, charginos} and {neutralinos}, ...~\cite{R,ssm,ff,fayet79, martin}.
They differ from ordinary particles by $1/2$ unit of spin and are distinguished by a  {\it $R$-parity\,}  quantum number related to baryon and lepton numbers, 
discrete remnant of a continuous $U(1)_R$ symmetry, 
making the lightest superpartner stable.

\vspace{1mm}

While the standard model \cite {ws,ws2} involves a single scalar doublet leading to one Brout-Englert-Higgs boson 
\cite{g,be,h,ghk}, 
spontaneous electroweak breaking is induced here
by {\it two doublets} $h_1$ and $h_2$. They are responsible for charged-lepton and down-quark masses, and 
up-quark masses, respectively, 
leading  to additional {\it charged and neutral\,\, \hbox{spin-0} BEH bosons}. 
These theories also provide systematic associations between massive gauge bosons and \hbox{spin-0} BEH bosons, 
a very non-trivial  feature owing to
{\it their different gauge symmetry properties}
 \cite{R,ssm,gh}.
 
\vspace{.8mm}
These relations were proposed in 1974 even before the standard model 
(SM) got considered as ``standard'', and are at the basis of its supersymmetric extensions,
even if they may often go unnoticed.
Weak neutral currents were just recently discovered \cite{nc} with their structure unknown, 
and the $W^\pm$ and $Z$ hypothetical.
Little attention was paid to fundamental \hbox{spin-0} particles, 
the very possibility of their existence getting questioned and frequently denied for many years later.

\vspace{1mm}

Proposing relations between massive spin-1 mediators of weak interactions and spin-0 particles 
associated with electroweak breaking and mass generation \cite{R,gh} then amounted to

\vspace{-4.5mm}
\be
\label{hyp}
\ba{c}
\hbox{\it relate two classes of hypothetical particles,}
\vspace{1mm}\\
\hbox{\it using an hypothetical symmetry\,!}
\ea
\ee

\vspace{-.5mm}
\noindent
And this at a time  when supersymmetry was viewed as an algebric structure 
\cite{gl,va,wz,ra}
very far from being able to describe nature, for many reasons including an obvious lack of similarities between known bosons and fermions.

\vspace{1mm}

Forty years later, the situation has improved considerably. With the introduction of $R$-odd superpartners 
and two spin-0 doublets for the electroweak breaking, supersymmetry could indeed be 
a symmetry of the fundamental laws of physics
\cite{ssm}.
The  discoveries in 1983 of the $W^\pm$ and $Z$ mediators of weak interactions \cite{w,z}, and 
in 2012 of a new boson considered as a spin-0 BEH boson \cite{higgs,higgs2} confirmed to a very large extent the validity of
the standard theory (SM)
or of a closely approaching one.
This gives additional interest to the relations 
between \hbox{spin-1} and spin-0 bosons provided by supersymmetric extensions of the standard model.

\vspace{1mm}
These relations may be more concretely discussed
now that we know, with the $Z$ and $h$ bosons, at least one representative in each class of formerly hypothetical particles. 
The Supersymmetric Standard Model offers a way to view the 125 GeV/$c^2$ boson recently observed at CERN 
as {\it a \hbox{spin-0} partner of the $Z$}, 
up to a mixing angle induced by supersymmetry breaking.

\vspace{-2mm}

  \section{The spin-0 \lowercase {\boldmath {$\,z\,$}} partner of the  $\,Z$}
\vspace{-2mm}

 Within supersymmetry two spin-0 doublets $h_1$ and $h_2$ are needed for the electroweak breaking, 
 at first to avoid a massless chiral chargino, allowing for the construction of {\it two Dirac winos} associated with 
 the $W^\pm$ within a massive gauge multiplet of supersymmetry \cite{R}.
  These doublets $h_1=(h_1^0,h_1^-)$ and $h_2=(h_2^+,h_2^0)$ have weak hypercharges $Y=-1$ and $+1$.
  By leading to a negative mass$^2$ term for $h_2$,
  the term $-\,\xi D'$  \cite{fi} associated with $U(1)_Y$ in the Lagran\-gian density plays a crucial role in triggering spontaneous electroweak breaking and giving masses 
  to the $W^\pm$ and $Z$, and to their spin-$\frac{1}{2}$ and spin-0 partners.
  
\vspace{1mm}
  
  The auxiliary components   {\boldmath $ D$}  and $D'$ 
  associated with the $SU(2)\times U(1)_Y$ gauge group are expressed as
\be
  \ba{ccc}
  \hbox{\,\boldmath $ D$}  \!&=&\!  -\,\hbox{\small $\dis \frac{g}{2}$}  \ (h_1^\dagger \, \hbox{\boldmath $ \tau$} h_1
 + h_2^\dagger \, \hbox{\boldmath $ \tau$} h_2)+\, ... \,,
 \vspace{.5mm}\\
 \,D' \!&=&  \xi +\,\hbox{\small $\dis \frac{g'}{2}$}  \ (h_1^\dagger \,h_1- h_2^\dagger \,h_2)+\, ...\ .
  \ea
  \ee
 The resulting potential reads  \cite{R}
 \be
\label{vpot}
\ba{c}
\!V =\, \hbox{\small $\dis \frac{1}{2}$}\ (\hbox{\boldmath$D$}^2 +D'^2)+ \,...\,= \, 
\hbox{\small $\dis \frac{g^2}{8}$}\ (h_1^\dagger \, \hbox{\boldmath $ \tau$} \,h_1+ h_2^\dagger \, \hbox{\boldmath $ \tau$} \,h_2+\,...\,) ^2 
\vspace{.5mm}\\
+ \hbox{\small $\dis \frac{1}{2}$}\ [\,\xi + \hbox{\small $\dis \frac{g'}{2}$}\,
(h_1^\dagger \,h_1- h_2^\dagger \,h_2)+\,...\,]^2+ \,... \ ,
\vspace{-5.5mm}\\
\ea
\ee

\vspace{3mm}
\noindent
ignoring for the moment possible soft  supersymmetry-breaking terms, considered at a later stage.
Its quartic part, fixed by the electroweak couplings $g$ and $g'$ as
 \be 
\label{quartic0}
V_{\,\rm quartic}\, = \  \frac{g^2+g'^2}{8}\ (h_1^\dagger \,h_1-h_2^\dagger\, h_2)^2 +\, 
\frac{g^2}{2}\  |h_1^\dagger \,h_2|^2\,,
\ee
appears within supersymmetry  as part of electroweak gauge interactions.

 \vspace{1.5mm}
 
The potential is minimum 
\vspace{-.3mm}
for $<\!h_1^0\!>\ =v_1/\sqrt 2\,$, \linebreak $<\!h_2^0\!>\ =v_2/\sqrt 2$\,.
The correspondence with the notations of \cite{R}, using two doublets $\varphi"$ and $\varphi'$ with the same $Y=-1$,
\,is as follows:
\be
\label{dic}
\ba{ccc}
h_1= \varphi"=\left(\ba{c} \!h_1^0=\varphi"^0 \vspace{.5mm}\\ h_1^-=\varphi"^-\ea\!\right),\ \ 
h_2=\left(\ba{c} \!h_2^+=-\varphi'^+ \vspace{.5mm}\\ h_2^0\,=\ \varphi'^{0*}\ea\!\right),
\vspace{2.5mm}\\
\tan\beta=v_2/v_1\  \equiv \  \tan\delta=v'/v" \,.
\vspace{-4.5mm}\\
 \ea
 \ee
 
 \vspace{4mm}

The $\mu \,H_1 H_2$ superpotential term is first taken to vanish,
as initially forbidden by a continuous $U(1)_R$ 
and/or an extra $U(1)_A$ symmetry  \cite{R}. 
The latter acts according to
$\varphi"\to e^{i\alpha}\, \varphi",\, \varphi'\to e^{-i\alpha}\, \varphi'$ as introduced 
in a pre-susy two-doublet model in \cite{2hd}, \,i.e.
\be
\label{uasym}
h_ 1\to \,e^{i\alpha}\, h_1\,,\ \ h_ 2\to \,e^{i\alpha}\, h_2\,,
\ee
allowing to rotate the phases of the two doublets independently.  Taking $\mu=0$ in this first stage
allows for gauge symmetry to be spontaneously broken with supersymmetry remaining conserved 
in the neutral sector, shedding light on the relations between massive gauge bosons and spin-0 BE-Higgs bosons provided by supersymmetry.

 \vspace{2mm}

The initial $U(1)_R$ symmetry survives the electroweak breaking induced by $<\!h_1\!>$ and $<\!h_2\!>$.
As long as it is present it allows us to benefit,
in the absence of a $\mu$ term and of direct gaugino mass terms,
from Dirac neutralinos as well as charginos.
And more specifically {\it two Dirac winos and a Dirac zino},  carrying $\pm 1 $ unit of the additive quantum number $R$.

 \vspace{1.5mm}
Some attention  may be useful in the presence of an extra $U(1)$ symmetry acting on $h_1$ and $h_2$ 
as in (\ref{uasym}) \cite{2hd}, that became known later as a $U(1)_{PQ}$ symmetry.
Indeed it might lead to a classically massless pseudoscalar $A$ (and associated scalar $s_A$), 
jointly described by \cite{R,ssm}
\be
\label{sa}
\varphi"^{0} \sin\delta+\varphi'^{0*}\cos\delta=\,h_1^0 \,\sin\beta + h_2^0 \,\cos\beta \,=
\frac{s_A+iA}{\sqrt 2}\,.
\ee
These particles, momentarily appearing as classically massless
in the spectrum \cite{ssm}, get a mass as in \cite{R} through an explicit breaking
of the $U(1)_A$ symmetry \footnote{Such a classically-massless or light pseudoscalar $A$ 
associated with a $U(1)_A$ symmetry is usually referred to as an axion or axionlike particle.}.

 \vspace{2mm}
 We see from (\ref{vpot}) that the term  $-\,\xi D'$ in $\cal L$  
generates  a negative mass$^2$ for $h_2$, 
 triggering spontaneous electro\-weak  breaking.
 The origin is a saddle point 
  \vspace{-.4mm}
 of the potential,
with 
$\,m^2(h_2)= - \frac{\xi g'}{2} < 0\,,\ m^2(h_1)= \frac{\xi g'}{2} > 0$.
The would-be spin-0 Goldstone field (with $\delta=\beta$ as indicated in (\ref{dic}))
\be
\label{zg}
\ba{ccc}
z_g\!&=&\! -\,\sqrt 2\ \,\hbox{\rm Im} \, (\varphi"^0\,\cos\delta  + \varphi'^0\,\sin\delta )
\vspace{1mm}\\
\!&=&\ \ \
\sqrt 2\ \,\hbox{\rm Im} \, (-h_1^0\,\cos\beta  + h_2^0\,\sin\beta )\,
\ea
\ee
is eliminated by the $Z$. The corresponding real part
\be
\label{z}
\ba{ccc}
\ z &=& \sqrt 2\ \,\hbox{\rm Re} \, (-\varphi"^0\,\cos\delta  + \varphi'^0\,\sin\delta )
\vspace{1mm}\\
\!&=&
\sqrt 2\ \,\hbox{\rm Re} \  (-h_1^0\ \cos\beta  + h_2^0\ \sin\beta )\,
\ea
\ee
describes {\it a scalar BEH boson associated with the $Z$ under supersymmetry}, 
with the same mass $m_Z$ as long as supersymmetry is unbroken \cite{R,ssm,gh}.

\vspace{2mm}

This results in the general association  
\be
\label{gh2}
\hbox {\framebox [8.55cm]{\rule[-.18cm]{0cm}{.65cm} $ \dis
Z\   \,\stackrel{SUSY}{\longleftrightarrow }\   \hbox{2 Maj. zinos} \ \,
\stackrel{SUSY}{\longleftrightarrow }\  \hbox{spin-0 BEH boson}\ z\,
$}}
\ee
with in this description
\be
\label{expz}
\hbox {\framebox [6.1cm]{\rule[-.18cm]{0cm}{.6cm} $ \dis
z= \sqrt 2\ \hbox{Re} \  (-\,h_1^0\,\cos\beta+h_2^0\,\sin\beta )\,.\,
$}}
\ee
This is also made possible by the $U(1)_R$ symmetry remaining unbroken at this stage,
allowing for the two Majorana zinos to combine into a {\it Dirac zino\,} of mass $m_Z$.
It implies the existence of a spin-0 BE-Higgs boson
of mass
\be
\hbox {\framebox [7.55cm]{\rule[-.18cm]{0cm}{.6cm} $ \dis
m \,\simeq \, 91\  \hbox{GeV}/c^2\ \ \hbox{\it up to susy-breaking effects}.
$}}
\ee
This result, valid   {\it independently of $\,\tan\beta$}, may now be
compared with the recent CERN discovery of a new boson with a mass close to  125 GeV$/c^2$ \cite{higgs,higgs2}.

\vspace{2mm}

The spin-0 field $z$  may also be compared with the SM-like BEH field
\be
\label{hsm}
h_{SM}= \sqrt 2\ \hbox{Re} \  (\,h_1^0\,\cos\beta+h_2^0\,\sin\beta )\,.
\ee
This $z$ 
has Yukawa couplings ``of the wrong sign'' 
to down quarks and charged leptons, acquiring their masses through Yukawa couplings to $h_1$ \cite{ssm}
\footnote{
This definition (\ref{z},\ref{expz}) of $z$  includes a change of sign as compared to 
\cite{R}, so that it behaves as $\sqrt 2$ Re $h_2^0$ for large $\tan\beta$. 
The other sign would give, equivalently, Yukawa couplings ``of the wrong sign'' 
to up quarks.
The choice (\ref{z},\ref{expz}) subsequently leads to
$D_Z= +\, m_Z\, z\, + \,... \,$ in (\ref{valdz2}), 
and to identify, from $D_Z=-\,m_Z^2\, C_Z +\,... \, $ in (\ref{dc}), 
 $\,C_Z$ 
 with $\,{-\,z}/{m_Z}+ ... \,$ within the massive gauge superfield (\ref{expsf}).}. 
It becomes very close to $h_{SM}$ at large $\tan \beta$, with
\be
\label{angle-1}
<\,h_{SM}\,|\ z>\ \ =\, -\,\cos 2\beta\,.
\ee

\vspace{1mm}

We thus  rotate neutral chiral superfields as indicated by (\ref{sa}-\ref{z}), according to
\be
\label{hzzz0}
\!\!\left\{ \ba{ccrcc}
H_z\!&=&\! \dis -\,H_1^0\,c_\beta  +H_2^0\,s_\beta\!&=&\! (z\,+i\,z_g)/\sqrt 2+\,... \ ,
\vspace{2mm}\\
H_A\!&=&\! \dis H_1^0 \,s_\beta + H_2^0 \,c_\beta \!&=&\! (s_A+iA)/\sqrt 2 +\,..\ .
\ea\right.
\ee
$H_z$ describes the would-be Goldstone field $z_g$ and \hbox{spin-0} $z$ associated with the  $Z$
as in (\ref{zg},\ref{expz}), while $H_A$ describes
the scalar and pseudoscalar
 \be
 \label{haa0}
 \left\{ \ba{ccc}
s_A\!&=&\!\sqrt 2\ \hbox{Re} \ ( h_1^0\,s_\beta+h_2^0\,\,c_\beta )\,,
\vspace{1.5mm}\\
A\!&=&\!\sqrt 2\ \hbox{Im} \ ( h_1^0\,s_\beta+h_2^0\,\,c_\beta )\,,
\ea \right.
\ee
discussed more later.

\section{\boldmath Electroweak breaking \vspace{1.5mm} and 
\hbox{\ \ $Z$ and {\lowercase{$\,z\,$}} masses}}

\vspace{-2.8mm}

From

\vspace{-3mm}

\be
\label{DD}
\left.
\ba{ccr}
D_3&=&\hbox{\small$\dis  \frac{g}{2}$}\ (-|h_1^0|^2+|h_2^0|^2)+ \,...\ ,
\vspace{1mm}\\
D'&=&\xi+\hbox{\small$\dis \frac{g'}{2}$}\  (|h_1^0|^2-|h_2^0|^2)+ \,...\ ,
\ea\right.
\ee
we get for $\,D_Z =D_3\,c_\theta - D'\,s_\theta, \ D_\gamma=D_3\,s_\theta+D'\,c_\theta \, $,
\be
\label{dzdg}
\!\!\!\left\{
\ba{ccl}
\vspace{-5mm}\\
D_Z\!
&=&\! - \,\xi\,s_\theta\,+ \hbox{\small$\dis \frac{\sqrt{g^2+g'^2}}{2}$}\  (|h_2^0|^2-|h_1^0|^2)+\,...\ ,
\vspace{1mm}\\
D_\gamma\!&=&
\ \ \ \xi\, c_\theta \  +\,0\,+\,\,...\ \,.
\vspace{1mm}\\
\ea
\right.
\ee

\noindent
We express $V= \frac{1}{2}\ (D_Z^2+D_\gamma^2) + ...\, $  as a function of $h_1^0$ and $h_2^0$ as
\be
V= \frac{1}{2} \, \hbox{\Large $\left(\right.$} \!
-\xi\,s_\theta + \hbox{\small$\dis \frac{\sqrt{g^2+g'^2}}{2}$}\  (|h_2^0|^2-|h_1^0|^2)+ ... \!
\hbox{\Large $\left.\right)$}^2\! +\,...\,.
\ee

\vspace{1mm}

Minimizing
this term fixes only
$ v_2^2-v_1^2$, leading to a flat direction associated with $s_A$. $ v_2^2-v_1^2$ adjusts so that
\be
\label{valdz}
\,<\!D_Z\!>\ =  - \,\xi\,s_\theta  + 
\hbox{\small$\dis \frac{\sqrt{g^2+g'^2}}{4}$}\  (v_2^2-v_1^2)\,\equiv\,\,0\,.
\ee
Expanding $D_Z$ in (\ref{dzdg}) at first order in $h_1^0$ and $h_2^0$ we have
\be
\ba{ccl}
\label{valdz2}
\!D_Z
\!&=&\! \,\frac{1}{2}\sqrt{g^2+g'^2}\ 
(-v_1\,\sqrt 2\ \hbox{Re} \, h_1^0+v_2\,\sqrt 2\ \hbox{Re} \, h_2^0)+\,...\,
\vspace{2mm}\\
\!&=&\! m_Z\ 
\sqrt 2\ \hbox{Re} \,  (-\,h_1^0\,c_\beta+h_2^0\,s_\beta )
\, +\, ...\, = m_Z\, z \, +\, ...\,,
\ea
\ee

\noindent
providing from 
$\,D_Z^2/2 = \frac{1}{2}\, m_Z^2\,z^2+ \,...\ 
$ the supersymmetric mass term for the spin-0 field $z$.

\vspace{2.5mm}

The parameter $\xi$ associated with $U(1)_Y$  determines $m_Z$, given by $m_Z^2= (g^2+g'^2)\,(v_1^2+v_2^2)/4\,$.
For $v_1\simeq 0$ we would get for $Z$ and $z$ (described by $\simeq \sqrt 2\  \hbox{Re} \,h_2$)
\be
m_Z^2=\,m_z^2\,\simeq \, - \,2\,m^2(h_2)=\,\xi g'\,,
\ee

\vspace{-2mm}

\noindent
i.e.

\vspace{-5mm}

\be
\label{zxi}
\hbox {\framebox [4.1cm]{\rule[-.20cm]{0cm}{.65cm} $ \dis
m_Z\,=\,m_z\simeq \,\sqrt {\,\xi\, g'}\,,
$}}
\ee

\vspace{-2mm}

\pagebreak

\noindent
up to radiative corrections, and supersymmetry-breaking effects for $m_z$.
With $g'=e/\cos\theta \simeq .345$ the $Z,\,W^\pm$ and spin-0 BEH boson masses get {\it fixed by the $\xi$ parameter\,} 
associated with $U(1)_Y$. This leads to
\be
\xi\,\approx \frac{m_Z^2}{g'}\,\simeq \,\frac{m_W \,m_Z}{e}\,\simeq \,2.4\ 10^4  \  \hbox{GeV}^2\,,
\ee
or equivalently
\be
\sqrt{\xi}\  \approx\ \frac{v}{2}\ \,\frac{1}{\sin\theta}\ \sqrt{\frac{e}{\cos\theta}}\ \approx \,150\ \hbox{GeV}\,,
\ee
up to radiative corrections.

\vspace{2mm}

{\it The $\xi$ parameter} \cite{fi} {\it determines here the $W^\pm\!$ and $Z$ masses}, a feature that may further 
persist
when $R$-odd squarks and sleptons acquire large mass$^2$, e.g. from the compactification of extra dimensions.
More generally we have 

\vspace{-6.5mm}

\be
\label{valxi}
 (-\cos 2\beta)\ m_Z^2 = \xi\,g' \,,
\ee
reducing  to (\ref{zxi}) for large $\tan\beta$.  
$\,\xi=0$ would be associated with $\tan\beta=1$, leaving at this stage $m_Z$ and $m_W$
unfixed, at the classical level \footnote{In a supersymmetric grand-unified theory 
with a semi-simple gauge group
$\xi$ must vanish at the GUT scale, requiring for two spin-0 doublets $v_1=v_2$ \,i.e.~$\tan\beta =1$ 
at this scale.}. In such a situation the scalar $s_A$ associated with this flat direction would describe a classically massless particle with dilatonlike couplings. 

\vspace{1.5mm}

We also have from (\ref{dzdg},\ref{valdz})

\vspace{-5mm}
\be
\label{dgamma}
<\!D_\gamma\!>\ = \xi\,c_\theta 
=  \frac{g}{4s_\theta}\  (v_2^2-v_1^2) = \,\frac{m_W^2}{e}\, (-\cos\,2\beta).
\ee
Having at this stage the photino as the Goldstone fermion
implies that charged particles only are sensitive 
to the spontaneous breaking of the supersymmetry.
Neutral ones remain mass degenerate
within massive ($Z$) or massless  ($\gamma$) multiplets of supersymmetry,
before the introduction of extra terms breaking the $U(1)_A$ and $U(1)_R$ symmetries, 
the latter reduced to $R$-parity.

\vspace{1.5mm}

We now discuss zinos, winos and charged spin-0 bosons within massive gauge multiplets, 
before returning to \hbox{spin-0} bosons, and how they may be described by 
{\it massive gauge superfields}, in contrast with the usual formalism.

\vspace{-1mm}

\section {Zinos and other neutralinos}

\vspace{-1.5mm}

The massive gauge multiplet of the $Z$ \cite{R} includes
a Dirac zino, 
obtained from chiral gaugino and higgsino components transforming 
under $U(1)_R$ according to
\be
\hbox{gaugino} \ \lambda_Z \to e^{\gamma_5\alpha}\ \lambda_Z,\ \ 
\hbox{higgsino} \ \tilde h_z\to e^{-\gamma_5\alpha}\ \tilde h_z\,.
\ee
It may be expressed as a massive Dirac zino  with $R=+1$,
\be
\label{expzino}
\lambda_{Z L}+ (- \tilde h_z)_R \,
= \,(\lambda_{3} c_\theta -\lambda's_\theta )_L+ (\tilde h_1^0\,c_\beta-\tilde h_2^0\,s_\beta )_{R}.
\ee
Or equivalently as two Majorana zinos, degenerate as long as $U(1)_R$ is preserved, 
with a mass matrix given in the corresponding  $2\times 2$ gaugino-higgsino basis by 
\footnote{If $\,\psi=a_L+b_R\,$ is a Dirac spinor constructed from two Majorana ones $a$ and $b$, 
its mass term may be expressed through a non-diagonal matrix, 
\vspace{.5mm}
as 
$-\,i\,m\,\bar\psi\psi=  -\,i\,m\,$ $(\,\bar b\, \frac{1-i\gamma_5}{2}\,a+\bar a \,\frac{1+i\gamma_5}{2}\,b\,)=  -\,i\,\frac{m}{2}\,(\bar a b + \bar b a)$.}
\be
\label{mzino}
{\cal M}_{\rm zinos} = \left( \ba {cc}
0 & m_Z
\vspace{1mm}\\
m_Z& 0
\ea\! \right).
\ee 
Supersymmetry remains unbroken in this sector, in the absence of direct gaugino ($m_1,m_2$) and higgsino ($\mu$) mass terms.

\begin{table}[tb]
\vspace{-2.5mm}
\caption{Minimal content of the Supersymmetric Standard Model (MSSM).
Gauginos $\,\lambda',\,\lambda_3$ mix with higgsinos
 $\,\tilde h_1^{\circ},\,  \tilde h_2^{\circ}$ into a photino, two zinos and a higgsino, further mixed into four neutralinos.\,\,\,The charged $w^\pm$ associated with $W^\pm$ is usually known as $H^\pm$. The scalars $(z,\,s_A)$ mix into 
 $h$ and $H$. \ \ The N/nMSSM also involves an extra singlet superfield $S$ with a trilinear superpotential 
 coupling $\,\lambda \,H_1H_2S$, leading 
 to an additional neutralino (singlino) and two singlet bosons.
\label{tab:mssm}}
\vspace{-1mm}
\begin{center}
\begin{tabular}{|c|c|c|} 
\hline 
&&\\ [-0.3true cm]
gluons       	 &gluinos ~$\tilde{g}$        &\\[-.5mm]
photon           &photino ~$\tilde{\gamma}$   &\\ [-1mm]
---------------&$- - - - - - - - -$&--------------------------- \\ [-4.5mm]
 

$\begin{array}{c}
\\ W^\pm\\ [0mm]Z \ \  \\ [0mm]
\\ \\
\end{array}$

&$\begin{array}{c}
\hbox {winos } \ \widetilde W_{1,2}^{\,\pm} \\ [0mm]
\,\hbox {zinos } \ \ \widetilde Z_{1,2} \\ [0mm]
\hbox {higgsino } \ \tilde h_A
\end{array}$

&$\left. \begin{array}{c}
w^\pm\\ [0mm]
 z \\
[0mm]
s_A, \,A
\end{array}\ \right\} 
\begin{array}{c} \hbox {BE-Higgs}\\ \hbox {bosons} \end{array}$  \\ &&\\ 
[-.6true cm]
\hline &&

\\ [-4mm]
&leptons ~$l$       &sleptons  ~$\tilde l$ \\[-.5mm]
&quarks ~$q$       &squarks   ~$\tilde q$\\ [-3.2mm]&&
\\ \hline
\end{tabular}
\ec
\vspace{-2mm}
\end{table}

\vspace{1mm}

This $2\times 2$ zino mass matrix  may be unpacked again into a $4 \times 4$  neutralino matrix expressed in the ($\lambda',\lambda_3,\,\tilde h_1^0,\,\tilde h_2^0)$ basis
using (\ref{expzino}).
Including additional $\Delta R=\pm 2 $ super\-symmetry-breaking contributions from gaugino ($m_1, m_2$) and higgsino ($\mu$)
mass terms
 it reads
  \be
{\cal M}_{\rm inos} =\left( \ba {cccc}
\ m_1 &\ 0 &      \!\!-s_\theta c_\beta \,m_Z & \ s_\theta s_\beta \,m_Z   
\vspace{1mm}\\
\ 0&\ m_2 &       \  c_\theta c_\beta \,m_Z & \!\!-c_\theta s_\beta \,m_Z
\vspace{1mm}\\
\!\!-s_\theta c_\beta  \,m_Z & \ c_\theta c_\beta  \,m_Z & \ \,0&-  \mu
\vspace{1mm}\\
\ s_\theta s_\beta \,m_Z&\!\!-c_\theta s_\beta  \,m_Z & - \mu & \ \,0
\ea \right).
\ee

\vspace{1mm}
\noindent
For equal gaugino masses $m_1=m_2$ the photino $ \lambda_\gamma = \lambda'\,c_\theta+\lambda_3\,s_\theta$ is  a mass eigenstate.
 The remaining $3\times 3$  mass matrix is expressed in the ($\lambda_Z,\tilde h_1^0,\,\tilde h_2^0)$ basis (with
  $ \lambda_Z = -\lambda'\,s_\theta+\lambda_3\,c_\theta$) as
  \be
\hbox{\small $\dis
\left( \ba {cccc}
\ m_2 &       \  c_\beta \,m_Z & -s_\beta \,m_Z
\vspace{.5mm}\\
 \  c_\beta  \,m_Z & \ \,0&-  \mu
\vspace{.5mm}\\
-s_\beta  \,m_Z & - \mu & \ \,0
\ea \right) 
$} \,,
\ee

\vspace{1mm}
\noindent
as seen from (\ref{expzino}). It
further simplifies for $\tan\beta =1$ into
\be
\hbox{\small $\dis
\left( \ba {cccc}
\ m_2 &       m_Z/\sqrt 2  & \,-\,m_Z/\sqrt 2 
\vspace{.5mm}\\
 \ \,  m_Z/\sqrt 2  &  0&\,-  \mu
\vspace{.5mm}\\
\!-\,m_Z/\sqrt 2  & - \mu &  \,0
\ea \right)
$}\,  .
\ee
\vspace{0mm}

\noindent
Next to a pure higgsino of mass $|\mu|$ corresponding to $(\gamma_5) \,(\tilde h_1^0+\tilde h_2^0)/\sqrt 2$, the two zinos constructed from  $\lambda_Z$ and $(\tilde h_1^0-\tilde h_2^0)/\sqrt 2\,$
have the mass matrix
\be
\label{mzino1}
{\cal M}_{\rm zinos} =
 \left( \ba {cc}
m_2 & m_Z
\vspace{1mm}\\
m_Z& \mu
\ea \!\right),
\ee 
as obtained directly from (\ref{mzino}).

\pagebreak
\vspace{1.5mm}

There may also be additional neutralinos, as described by the extra N/nMSSM singlet $S$ 
with a $\,\lambda\, H_1 H_2 S\,$ superpotential coupling,
\vspace{-1mm}
leading through $\,<\!H_i\!>\ = \frac{v_i}{\sqrt 2}$ 
to a $\,\frac{\lambda v}{\sqrt 2}\ H_A S$  superpotential mass term. 
\vspace{-.7mm}
Here
$ H_A=H_1^0 \,s_\beta+ H_2^0 \,c_\beta $ is the same ``left-over''
chiral superfield as obtained in  (\ref{hzzz0}), now acquiring a mass by combining with $S$ \cite{R}.

\section{\boldmath The spin-0 \lowercase {\boldmath {$\,w^\pm\,$}} partner \vspace{1.5mm}
of the $\,W^\pm$, \hbox{\ \ and associated winos}}

\vspace{-2mm}

 We have, in a similar way,
\be
\label{gh3bis}
\hbox {\framebox [8.5cm]{\rule[-.18cm]{0cm}{.65cm} $ \dis
W^\pm\ \stackrel{SUSY}{\longleftrightarrow }\ \hbox{2 Dirac winos} \ \stackrel{SUSY}{\longleftrightarrow } \ 
\hbox{
spin-0 boson} \ w^\pm,
$}}
\ee
with 
$m_{w^\pm}\!=m_{W^\pm}$, 
also up to supersym\-metry-breaking effects. This is why the charged boson now known as $H^\pm$
was  called $w^\pm$ in \cite{R}.

\vspace{2mm}

The two doublets being expressed 
\vspace{-.3mm}
as $\,\varphi"\!=(h_1^0, h_1^-)$ and $\,\varphi'=(h_2^{0\,*},-h_2^-$)), 
as in (\ref{dic}) with $\delta= \beta$,
the would-be Goldstone field
\be
\label{wg}
w^\pm_g=  \varphi"^\pm\cos\delta+\varphi'^\pm\sin\delta=
h_1^\pm\cos\beta-h^\pm_2\sin\beta
\ee
is eliminated by the $W^\pm$. 
The orthogonal combination
\be
\label{w}
w^\pm\,= \varphi"^\pm\sin\delta-\varphi'^\pm\cos\delta =  h_1^\pm\sin\beta+h_2^\pm \cos\beta\,
\ee
(approaching  $\,h_1^\pm$ at large $\tan\beta$)
describes {\it a charged \hbox{spin-0} BEH boson associated with the $W^\pm$}
\cite{R,ssm,gh}.
\vspace{1.5mm}

With 
\be
\ba{ccl}
h_1^\dagger h_2= h_1^{0*} h_2^++ h_1^+ h_2^0 \!&=&\! 
\hbox{\small$\dis \frac{v}{\sqrt 2}$}\ (h_1^+\sin\beta+h_2^+\cos\beta)
+ ... 
\vspace{1mm}\\
\!&=&\! \hbox{\small$\dis \frac{v}{\sqrt 2}$}\ \,w^+\,+\, ...\ ,
\vspace{-6mm}\\
\ea
\ee

\vspace{3mm}
\noindent
the quartic terms (\ref{quartic0}) in the potential,
\be
V=\hbox{\small$\dis\frac{g^2}{2}$}\ (h_1^\dagger h_2)^2\! + \,...\,=\,
\hbox{\small$\dis\frac{g^2v^2}{4}$} \ 
|w^+|^2 + ...\,,
\ee
generate a mass $m_w=m_W={gv}/{2}$ \,for 
\be
\label{expw}
\hbox {\framebox [6cm]{\rule[-.2cm]{0cm}{.65cm} $ \dis
w^\pm\,\equiv H^\pm = h_1^\pm\sin\beta+h_2^\pm\cos\beta\,.
$}}
\ee
It is
the same as for the $W^\pm$, to which it is related by {\it two\,} infinitesimal supersymmetry transformations.

\vspace{2mm}

The mass spectrum is given, at this first stage for which supersymmetry is spontaneously broken
with the photino as the Goldstone fermion, by \cite{R}

\vspace{-4mm}

\be
\label{mwinos}
\left\{
\ba{cc}
m_{w^\pm}^2\,=\ m_{W^\pm}^2= \ \hbox{\small$\dis \frac{g^2\,(v_1^2+v_2^2)}{4}$}\,,
\vspace{1mm}\\
m^2({\hbox{winos}_{1,2}})\, =\ \hbox{\small$\dis\frac{g^ 2 v_{1,2}^ 2}{2}$}\ =\,  m_W^2\, (1\,\pm\,\cos\,2\beta)
\vspace{1mm}\\
\ \ \ \ \ \ \ \,=\ m_W^2\,\mp\ e\, <\!D_\gamma\!>\,.
\vspace{2mm}\\
\ea\right.
\ee

\vspace{-2mm}
\noindent
Boson-fermion mass$^2$ splittings are given by \hbox{$\,\pm \,e  <\!D_\gamma\!>\,$} 
(as in \cite{fayet79} in the absence of other sources of supersymmetry breaking),
and fixed by (\ref{dgamma}).

\vspace{2mm}

The two Dirac winos are  $R$ eigenstates carrying $R=\pm 1$, with masses   $gv_1/\sqrt 2$ and $gv_2/\sqrt 2$. 
The wino mass matrix would be supersymmetric (as for zinos in (\ref{mzino})) 
 for $\xi=0$ so that $\beta=\pi/4$ 
and  m(winos) $= m_W$, with $<\!D_\gamma\!>\ =\ <\!D_Z\!>\  = \, 0$ \ from (\ref{dgamma}).
\vspace{2mm}

In the presence of additional $\Delta R = \pm 2\,$ gaugino and higgsino mass terms  further breaking the supersymmetry
as well as $U(1)_R$ (for $m_2$ and $\mu$) and  $U(1)_A$  (for $\mu$),
the wino mass matrix obtained from (\ref{mwinos}) reads
\be
\label{mwinos1}
 \hbox{$\dis
{\cal M}_{\rm winos} \,= \left(\ba {cc}
m_2 \!\!& \!\!\!\!\hbox{\small$\dis\frac{gv_2}{\sqrt 2}$}=m_W\sqrt 2\, s_\beta
\vspace{1mm}\\
\!\hbox{\small$\dis\frac{gv_1}{\sqrt 2}$}=m_W\sqrt 2 \, c_\beta \!\!& \!\!\!\!\!\mu
\ea\! \right).
$}
\ee 
$m_2$ and $\mu$ jointly allow for both winos to be heavier than $m_W$ (as experimentally required \cite{pdg}).

\vspace{2mm}

 For  gaugino and higgsino mass terms related by $m_1=m_2=m_3=-\mu$ (up to radiative corrections), 
 possibly also equal to the gravitino mass $m_{3/2}$, with $\tan\beta=1$ \cite{mwinos},
 we get from (\ref{mzino1},\ref{mwinos1}) remarkable mass relations like, at the classical level,
\be
\label{winos}
 \left\{\
 \ba{ccc}
 m^2(\hbox{winos}) \!&=&\!m_W^2+m_{3/2}^2\,,
 \vspace{.5mm}\\
 m^2(\hbox{zinos})  \!&=&\!m_Z^2 +m_{3/2}^2\,,
  \vspace{.5mm}\\
  m(\hbox{photino})  \!&=&\!  m(\hbox{gluinos})\,=\,m_{3/2}\,.
 \ea\right.
 \ee
 This also paves the way for more general situations involving \hbox{$N=2$} extended supersymmetry 
 with grand-unification groups \cite{gut,gutbis}.
 Similar mass relations like
 \be
 \label{xinos}
 \!\!\left\{\,
 \ba{ccl}
 m^2(\hbox{xinos}) \!&=&\!m_X^2+m_{3/2}^2\,,
 \vspace{.5mm}\\
 m^2(\hbox{yinos})  \!&=&\!m_Y^2 +m_{3/2}^2= m_X^2 +m_W^2+ m_{3/2}^2 \,,\!\!
 \ea\right.
 \ee
 are then  obtained for xinos, yinos, etc., with a grand-unification gauge group like $SU(5)$ or $O(10),\, ...\ $.
 
\vspace{2mm}
 
 Extra compact dimensions may then be responsible for supersymmetry and grand-unification breakings 
 \cite{gutbis}, 
 $R$-odd  supersymmetric particles carrying momenta $\pm m_{3/2}$ along an extra dimension. 
 When $R$-parity is identified with the action of performing a closed loop along such a compact dimension, 
 $m_{3/2}$ 
 and more generally superpartner masses
 get quantized in terms of its size, according to (\ref{winos},\ref{xinos})
 with  e.g.~in the simplest case
 \be
 m_{3/2}\,=\,(2n+1)\ \,\frac{\pi\hbar}{Lc}\,=\,(2n+1)\ \frac{\hbar}{2Rc}\ .
 \ee
 But let us return, in a more conservative way, to 4 spacetime dimensions.
 
\vspace{-1mm}

\pagebreak

\section{\boldmath \hspace{-1mm}The pseudoscalar $\,A\,$ and scalar $\,s_A$}

\vspace{-.5mm}

The potential  (\ref{vpot}) admits, at this initial stage excluding a $\mu \,H_1H_2$ superpotential term (both 
$U(1)_R$ and  $U(1)_A$ symmetries being present)
two clas\-sically flat directions corresponding to the scalar $s_A$ and pseudoscalar $A$ in (\ref{sa}), both classically massless \cite{ssm}.
 $A$ then appears as an ``axion'' associated with 
the extra $U(1)_A$ symmetry acting on $h_1$ and $h_2$ as in (\ref{uasym}) \cite{2hd},
extended to supersymmetry  according to \cite{R}
\be
\label{ua0}
H_{1}\ \to \ e^{i\alpha} \, H_1\,,\ \ 
H_{2}\ \to \ e^{i\alpha} \, H_{2}\,.
\ee 
Its scalar partner $s_A$ is also associated with a flat direction,
the minimisation of the potential (\ref{vpot}) fixing only $v_2^2-v_1^2$.

\vspace{2.5mm}

This ``axion'' $A$ (a notion unknown at the time, that appeared in a different context several years later) and associated scalar $s_A$ 
were given a mass in \cite{R} by breaking explicitly the $U(1)_A$ symmetry (\ref{uasym},\ref{ua0}), 
now often referred to as $U(1)_{PQ}$. This was done by introducing a singlet $S$ coupled 
through a trilinear superpotential 
$\lambda\,H_1 H_2 S$, and transforming under 
$U(1)_A$ according to 
\be
S \to e^{-2i\alpha}\,S\,.
\ee
Its $f(S)$ superpotential interactions, 
that may include $S,\,S^2$ and $S^3$ terms as in the N/nMSSM, break explicitly $U(1)_A$, 
the presence of a quasimassless ``axion'' being avoided.

\vspace{2.5mm}

Explicitly, the potential includes an extra term $V_\lambda$, with a vanishing minimum
still preserving the supersymmetry. It reads
\be
\ba{ccl}
\!V_\lambda\!\!&=&\! \hbox{\small$\dis\left|\frac{\partial{\cal W}}{\partial S}\right|$}^2
=\,|\,\lambda \,h_1 h_2 +\sigma + ...\,|^2 + ...\,
\vspace{2mm}\\
\!&=&\!\! 
\hbox{\small$\dis \frac{\lambda^2v^2}{2}$}\,|\underbrace{h_1 s_\beta + h_2 c_\beta}_{h_A}|^2 
+ ... =  \hbox{\small$\dis \frac{1}{2}$} 
\hbox{\small$\dis \frac{\lambda^2v^2}{2}$}\, (s_A^2+A^2) 
+ ... \,.
\vspace{-4mm}\\
\ea
\ee

\vspace{1mm}
\noindent
It provides a mass term ($\lambda v/\sqrt 2$) for the complex field
\be
h_A=\,\hbox{\small$\dis\frac{s_A+iA}{\sqrt 2}$}=\ h_1 s_\beta + h_2 c_\beta\,,
\ee
the would-be ``axion'' $A\,$ (and associated scalar $s_A$) acquiring a mass 
$m_A=\lambda v/\sqrt 2\,$ \cite{R}.

\vspace{2mm}

In terms of superfields, the $\lambda\,H_1 H_2\,S$ superpotential coupling of the N/nMSSM generates in \cite{R}, 
from $\ <H_1>\ = {v_1}/{\sqrt 2}\,,\ <H_2>\ = {v_2}/{\sqrt 2}\,$,
\be
\lambda\,H_1 H_2\,S \,=\,\hbox{\small$\dis\frac{\lambda \,v}{\sqrt 2}$}\ (H_1 s_\beta+H_2 c_\beta)\ S\,+\,...
\,=\,\hbox{\small$\dis\frac{\lambda \,v}{\sqrt 2}$}\ H_A S\,+\,...\,,
\ee
a supersymmetric mass term $\lambda v/\sqrt 2$ for $H_A$ and $S$,
possibly to be combined  with a $\frac{1}{2}\, \mu_S S^2$ singlet mass term, if present.

\section{\boldmath {\lowercase {$\,z$}} \,Yukawa couplings \vspace{1mm} ``of the wrong sign''}

\vspace{-.5mm}

The new boson found at CERN close to 125 GeV/$c^2$
\cite{higgs,higgs2} 
is considered as a Brout-Englert-Higgs boson \cite{g,be,h,ghk} associated with the electroweak breaking, 
as expected in the standard model \cite{ws,ws2} where this breaking involves a single spin-0 doublet.
But it may also be interpreted, in general up to a mixing angle, as a spin-0 partner of the $Z$ 
under {\it two} infinitesimal supersymmetry transformations.
The $z$ field in (\ref{expz}) may be compared 
with the SM-like scalar, obtained from the real part of the neutral component of ``active'' doublet combination
\be
\varphi_{\rm sm}=\varphi" \cos\delta +\,\varphi'\sin\delta\,=h_1 \cos\beta +\,h_2^c \,\sin\beta\,,
\ee
 such that
$<\varphi_{\rm sm}^0>\ =\frac{v}{\sqrt 2}$, \,and
\be
h_{SM}=\, \sqrt 2\ \,\hbox{\rm Re} \, (\,h_1^0\,\cos\beta  + \,h_2^0\,\sin\beta \, )
\ee
as in (\ref{hsm}).
We have $\,<h_{SM}\,|\ z>\  = -\,\cos 2\beta$,
the two fields
getting close for large $\tan\beta$, with the $z$ tending 
to behave very much as the SM-like 
$\,h_{SM}$.

\vspace{1.5mm}
More precisely 
\vspace{-.4mm}
while $h_{SM}$ has standard Yukawa couplings to quarks and charged leptons 
$\,m_{q,l}/{v}= 2^{1/4}\,G_F^{1/2}$ $m_{q,l}$,
the $z$ has almost-identical couplings
\be
\label{zcoup}
\frac{m_{q,l}}{v}\ 2\,T_{3\,q,l}\,= \ 2^{1/4}\,G_F^{1/2}\ m_{q,l}\ 2\,T_{3\,q,l}\,.
\ee
They simply differ by {\it a relative change of sign for $d$ quarks and charged leptons} (with $2\,T_{3\,d,l}\!=-1$) acquiring their masses through $<h_1^0>$, 
as compared to $u$ quarks.

\vspace{1.5mm}

This may also be understood by {\it deducing the scalar couplings of the spin-0 $z$ 
from the axial couplings of the spin-1 $Z$}, as follows:

\vspace{1mm}

The $Z$ is coupled,  with coupling $\sqrt{g^2+g'^2}$,  to the weak neutral current $J_Z^\mu=J_3^\mu-\sin^2 \theta \,J^\mu_{\rm em}$,
with an {\it axial part} $J^\mu_{3\ \rm ax}$ fixed by $T_{3\,q,l}/2$. It gets its mass by eliminating  the Goldstone field $z_g$, pseudoscalar partner of the scalar $z$. 
 As seen from the global limit $g,\,g'\to 0$ for which the $Z$ would become massless and behave like the spin-0 $z_g$,  this $z_g$ has
{\it pseudoscalar} couplings to quarks and leptons given by
\be
\label{zu}
\ba{l}
\sqrt{g^2+g'^2}\ \,\hbox{\small $\dis \frac{T_{3\,q,l}}{2}\ \,\frac{2m_{q,l}}{m_Z}$}\ =
\vspace{2mm}\\
\ \ \ \  \ =\ \hbox{\small $\dis  \frac{m_{q,l}}{v}\ 2\,T_{3\,q,l}$}\,= \ 2^{1/4}\,G_F^{1/2}\ m_{q,l}\ 2\,T_{3\,q,l}\,.
\ea
\ee
This is the same argument as for relating the axial coupling of a $U$ boson to the pseudoscalar coupling
of the equivalent axionlike pseudoscalar $A$ or $a$, with the $U$, replaced by the $Z$, considered
in the small mass and small coupling limit \cite{U}.
The scalar partner $z$, described by the same chiral superfield $H_z$ 
as the would-be Goldstone $z_g$, has {\it scalar} couplings to quarks and leptons 
also given by (\ref{zu}) (and as found in (\ref{zcoup}) by the conventional method).
This may be remembered as 
\be
\label{zu1}
\ba{l}
\hbox {\framebox [8.5cm]{\rule[-.7cm]{0cm}{1.65cm} $ \dis
\ba{c}
\underbrace{\sqrt{g^2+g'^2}\ \,\hbox{\small $\dis \frac{T_{3\,q,l}}{2}$}}_{\hbox{\small axial coupling of $Z$}}\ \,\hbox{\small $\dis\frac{2m_{q,l}}{m_Z}$}
\ = \ \underbrace{2^{1/4}\,G_F^{1/2}\ m_{q,l}\ 2\,T_{3\,q,l}}_{\hbox{\small scalar coupling of $z$}}\,.
\vspace{-5mm}\\
\leftrightarrow\ \ \ \ 
\ea
$}}
\ea
\ee

\vspace{1.5mm}
This also provides the couplings of the spin-0 $w^\pm$ 
from the $W^\pm$ ones using (\ref{wg},\ref{w}), leading to the factors 
$m_{d,e}\tan\beta$ and $m_u\cot \beta$ in the expressions of these couplings.

\vspace{1.5mm}

We recover as expected spin-0 couplings to quarks and leptons proportional to their masses, 
in contrast with the couplings of the 
\hbox{spin-1} $Z$ and $W^\pm$ in
the same multiplets of supersymmetry. 
This is, however, a rather intriguing feature as $z$ and $Z$, or $w^\pm$ and $W^\pm$ 
may also be simultaneously described by the same massive gauge superfields
$Z(x,\theta,\bar \theta)$ and $W^\pm(x,\theta,\bar \theta)$. 
It is discussed and understood in \cite{gh},  showing how the couplings 
of the spin-0 $z$ and $w^\pm$ get in this description {\it resurrected}
from the supersymmetric mass terms for quarks and leptons, through non-polynomial field and superfield redefinitions.

\vspace{1.5mm}

In comparison with a standard model $h_{SM}$ boson the $z$ has {\it reduced trilinear couplings to the $W^\pm$ and $Z$} by a factor $-\cos2\beta$
owing to (\ref{hsm},\ref{angle-1}),
so that
\be
\!\left\{\! \!\ba{ccl}
(z\,VV)\,   \hbox{\small couplings}\!&=&\! (h_{SM}\,VV)\ \hbox{\small couplings} \times  (-\cos 2\beta),
\vspace{1.5mm}\\
(z\,ff)\,   \hbox{\small couplings}\!&=&\! (h_{SM}\,VV)\, \hbox{\small couplings}  \times (2T_{3f}=\pm 1).
\ea\right.
\ee

\noindent
The expected production of a $z$ in the $ZZ^{*}$  or $WW^{*}$ decay channels would then be {\it decreased by $\cos^2 2\beta$}
as compared to a SM boson, with respect to fermionic quark and lepton channels (the change of sign in $d$-quark and charged-lepton couplings 
also affecting the $h\to\gamma\gamma$ decay).

\vspace{1.5mm}

But the $z$ does not necessarily correspond to a mass eigenstate, and 
further mixing effets induced by supersymmetry breaking must be taken into account, as discussed in Sec.~\ref{sec:Nnmssm}.
The $h$ field presumably associated 
with the 125 GeV$/c^2$ boson observed at CERN may then be expressed (in the absence of further mixings effects that could involve an additional singlet) as
\be
h= \sqrt 2 \ \hbox{\rm Re} \,( -h_1^0 \,c_{\beta'} +\,h_2^0 \,s_{\beta'}) = \sqrt 2 \ \hbox{\rm Re} \,( -h_1^0 \,s_\alpha +\,h_2^0 \,c_{\alpha})
\ee
with $\beta'=\frac{\pi}{2}-\alpha$, \,and 
\be
<z\,|\,h>\ = \,\cos \,(\beta-\beta'\,)\,=\,\sin\,(\beta+\alpha)\,,
\ee
At the same time 
\be
<h_{SM}\,|\,h>\ = \,-\cos \,(\beta+\beta'\,)\,=\,\sin\,(\beta-\alpha)\,,
\ee
the factor $\cos^2 2\beta$ affecting the $ZZ^*$ or $W W^*$ decay rates of a $z$  being replaced by
$\cos^2 (\beta+\beta')=\sin^2(\beta-\alpha).$
The physical mass eigenstate $h$ is very close to the $z$ in (\ref{expz})
for $\,\beta=\beta'$ \,i.e. $\beta+\alpha \simeq\frac{\pi}{2}$, \,then justifying {\it an almost  complete association of this 125 GeV$/c^2$ boson with the spin-1 $Z$}.

\section{Massive \vspace{1.5mm}   gauge superfields 
\hbox{for spin-0 bosons}}

\vspace{-2mm}

 Supersymmetric theories thus allow for 
associating spin-1 with spin-0 particles within massive gauge multiplets of supersymmetry,  
leading to {\it gauge/BE-Higgs unification}, BEH bosons appearing as 
{\it extra spin-0 states of massive spin-1 gauge bosons}.
We can even use the superfield formalism \cite{sf} to jointly describe these massive \hbox{spin-1}, 
spin-$\frac{1}{2}$ and now also spin-0 particles  
with {\it massive gauge superfields}  \cite{gh}. 

\vspace{1mm}
Quite remarkably, this is possible  {\it in spite of their different electroweak properties}, spin-1 fields 
transforming as a gauge triplet and a singlet with spin-0 BEH fields transforming as electroweak doublets.
And although gauge and BE-Higgs bosons have {\it very different couplings to quarks and leptons},
which may first appear very puzzling but is elucidated in \cite{gh}, using appropriate
changes of field and superfield variables.
\vspace{1mm}

To do so we must {\it change picture} in our representation of such spin-0 bosons. The previous $z$ and $w^\pm$ ($\equiv H^\pm$) 
cease being described by spin-0 components of the chiral superfields $H_1$ and $H_2$, to get described,
through a {\it non-polynomial change of (super)fields}, by the lowest ($C$) components of the $Z$ and  $W^\pm$ superfields.
This association 
can be realized in a supersymmetric way 
by {\it completely\,} gauging away the three chiral superfields $H_1^-,\,H_2^+$ 
and $H_z = -\,H_1^0\, c_\beta+H_2^0\, s_\beta$. These complete superfields are now considered as Goldstone chiral superfields 
and eliminated by being taken identical to their v.e.v.'s:
\be
\label{chirgold}
\left\{ \,
\ba{l} 
\ \ \ 
H_1^- \equiv \,H_2^+\equiv \,0\,,
\vspace{1mm}\\
H_z = -\,H_1^0\, c_\beta+H_2^0\, s_\beta\ \equiv\ \,-\,\hbox{\small $\dis \frac{v}{\sqrt 2}$}\,\cos 2\beta\ .
\ea \right.
\ee
The field degrees of freedom normally described by them, 
i.e.~the spin-0 BEH fields referred to as $z$ and $w^\pm$ in (\ref{expz},\ref{expw}) 
and associated higgsino fields
are completely gauged away, and naively seem to be ``lost'' in this description.

\vspace{1mm}
But at the same time the corresponding gauge superfields $Z(x,\theta,\bar\theta)$ and 
$W^\pm(x,\theta,\bar\theta)$ 
acquire masses in a supersymmetric way,  describing new physical degrees of freedom. These correspond precisely to those 
just ``lost'' in the gauging-away of
$H_1^-,H_2^+$ and $H_z$ in (\ref{chirgold}).
The chiral superfields, 
\vspace{-.3mm}
normalized so that  $\,<\!H_i^0\!>\ =\ <\!h_i^0\!>\  =$\linebreak $v_i/\sqrt 2\,$,
generate mass terms 
$\,\frac{1}{2}\,m_Z^2 \,(Z^2)_{D} $ and $\,m_W^2$\linebreak  $ |W^+|^2_{\,D} $  for $Z(x,\theta,\bar\theta)$ and $W^\pm(x,\theta,\bar\theta)$,
the linear term in $Z(x,\theta,\bar\theta)$ vanishing owing to (\ref{valdz}).

\vspace{2mm}

In the superfield formalism for supersymmetric gauge theories \cite{sf,wz2,ym} the Lagrangian density \cite{R} 
includes the terms 
\be
\label{lag}
\ba{l}
\!\!\!{\cal L}\,  = \, \frac{1}{2}\, \left[\,H_1^\dagger \,\exp(g \,\hbox{\boldmath $ \tau. W$}-g' B)\,H_1 \right.\ \ \ \ \ \ \ \ \ \ \ \ \ \ \ \ 
\vspace{.3mm}\\
\ \ \ \ \ \ \ \ \left. + \ H_2^\dagger\, \exp(g \,\hbox{\boldmath $ \tau. W$}+g' B)\,H_2\, \right]_D \!  -\, \xi\,D' + \,...\,.\!\!\!\!
\ea
\ee
We make the generalized gauge choice (\ref{chirgold}), so that 
\be
H_1=\left(\!\ba{c}\frac{v_1}{\sqrt 2} + ...\\ 0 \ea\!\right),\ \ \ H_2=\left(\!\ba{c}0 \\ \frac{v_2}{\sqrt 2}+ ...\ea\!\right),
\ee
 the ... involving the left-over superfield $H_A$.
A second order expansion of $\cal L$  along the lines of \cite{gh}, with 
\be
-\xi \,D'\,=\, \xi\ \sin\theta \, D_Z-\,\xi\ \cos\theta \,D_\gamma\,.
\ee
generates superfield  mass terms for $W^\pm(x,\theta,\bar\theta)$ and $Z(x,\theta,\bar\theta)$. 
The term linear in $Z(x,\theta,\bar\theta)$, which appears  with the coefficient
\be
\frac{ \sqrt{g^2+g'^2}}{4}\ (v_1^2-v_2^2)+ \xi\sin\theta\,=\,-\ <D_Z>\ \equiv \,0\,,
\ee
vanishes identically  owing to (\ref{valdz}).

\vspace{1.5mm}

We get at second order 
\be
\ba{ccc}
{\cal L}&=&\hbox{\small$\dis \frac{1}{2}$}\ \left[\,\hbox{\small$\dis \frac{(g^2+g'^2)(v_1^2+v_2^2)}{4}$}\ (W_3\cos\theta-B\sin\theta)^2\right.
\ \ \ \ \ \ \ \  \
\vspace{1mm}\\
 && \left.+ \, \hbox{\small$\dis \frac{g^2(v_1^2+v_2^2)}{4}$}\ (W_1^2+W_2^2)\right]_D + \,...\,,\vspace{-12mm} \\
\ea
\ee

\vspace*{6mm}

\noindent
so that
\be
\ba{ccl}
{\cal L}&=& \hbox{\small$\dis \frac{1}{2}$}\ m_Z^2\  (Z^2)_{D}\, + \,m_W^2\ |W^+|^2_{\ D}\,  + \,...
\vspace{2mm} \\
&=&\hbox{\small$\dis \frac{1}{2}$}\ m_Z^2\  (2\,C_Z D_Z -\partial_\mu C_Z\,\partial^\mu C_Z -Z_\mu Z^\mu + ...\,)
\vspace{.5mm} \\
&&\ \ \ -\,\hbox{\small$\dis \frac{1}{4}$}\ Z_{\mu\nu} Z^{\mu\nu} + ...\,+ \,\hbox{\small$\dis \frac{D_Z^2}{2}$}+\, ...\,+\, ...\ .
\vspace{-5.1mm} \\
\ea
\ee

 \vspace{3mm}
 \noindent
After elimination of auxiliary fields through
\be
\label{dc}
D_Z\,=\,-\,m_Z^2\ C_Z + ...\,= \,m_Z\,z+ ...\, ,\ \ \ \hbox{etc.}\,,
\ee
it includes the kinetic and mass terms 
for the gauge boson $Z$ and associated 
spin-0 boson $z$, 
\be
{\cal L}= -\,\hbox{\small$\dis \frac{1}{4}$}\, Z_{\mu\nu} Z^{\mu\nu} -\hbox{\small$\dis \frac{m_Z^2}{2}$}\ Z_\mu Z^\mu 
- \hbox{\small$\dis \frac{1}{2}$}\,\partial_\mu z\,\partial^\mu z -
\hbox{\small$\dis \frac{m_Z^2}{2}$}\ z^2 +\, ...\,.
\ee
And similarly for the $W^\pm$ and spin-0 partner $w^\pm$ \hbox{($\equiv H^\pm$)},
keeping also in mind that supersymmetry is spontaneously broken for this superfield when $\tan\beta\neq 1$.

 \vspace{2mm}
 
In this picture these spin-0 bosons 
get described by the lowest ($C$) spin-0 components of {\it massive} $Z$ and  $W^\pm $ superfields, expanded as
$\,Z(x,\theta,\bar\theta)=C_Z + ...  -\,\theta\sigma_\mu \bar\theta\   Z^\mu + \,...\,,\ 
W^\pm(x,\theta,\bar\theta)=C_W^\pm +\, ... \, -\,\theta\sigma_\mu \bar\theta\ W^{\mu\,\pm } +\, ...\ .$
Their $C$ components now describe,
through {\it non-polyno\-mial field transformations}\,
linearized as $
\,z=-\,m_Z \,C_Z+\,...\ $, $w^\pm = \,m_W\,C_W^{\pm} +\,... \,$, the same spin-0 fields $z$ and $w^\pm$ 
as in the usual formalism (with signs depending on previous choices for the definitions of $z$ and $w^\pm$). 
We thus have
\be
\label{expsf}
\hbox {\framebox [8.5cm]{\rule[-.65cm]{0cm}{1.6cm} $ \dis
\ba{ccl}
Z(x,\theta,\bar\theta)\!\!&=&\!\, (\,\hbox{\small$\dis\frac{-z}{m_Z}$}+ ...\,)\ +\, ... \,-\,\theta\sigma_\mu \bar\theta\    Z^\mu\, + \,...\, , 
\vspace{2mm}\\
W^\pm(x,\theta,\bar\theta)\!\!&=& \!\! (\,\hbox{\small$\dis\frac{w^\pm}{m_W}$}\!+ ...) +\,... 
-\,\theta\sigma_\mu \bar\theta\      W^{\mu\,\pm}\! + ...\, ,
\ea
$}}
\ee
{\it massive gauge superfields now describing spin-0 fields usually known as BEH fields\,!}
Their subcanonical ($\chi$)  spin-$\frac{1}{2}$ components, instead of being gauged-away as usual, now 
also correspond to physical degrees of freedom
describing
the \hbox{spin-$\frac{1}{2}$} fields previously known as higgsinos.

\section{\boldmath The BE-Higgs boson \vspace{1.3mm} as spin-0 partner of $Z$, in the \,(N/\hbox{\lowercase {n}})MSSM}
\label{sec:Nnmssm}

\vspace{-1mm}

\subsection{MSSM}

\vspace{-2mm}

This applies to the spin-0 sector of the MSSM. The scalar potential may be expressed by adding 
to $V$ obtained from (\ref{vpot}) 
the soft dimension-2 supersymme\-try-breaking term
\be
\label{ma0}
-\,m_A^2\ \,| h_1 s_\beta -\,h_2^c c_\beta|^2= -\,m_A^2\ \,|\varphi_{\rm in}|^2\,
\ee
including 
\vspace{-.4mm}
in particular the $\mu$-term contribution.
This term, which  vanishes for  $\, <h_i^0>\ =v_i/\sqrt 2$,
is a mass term for the doublet $\varphi_{\rm in}$, which has no v.e.v.~and thus no direct trilinear couplings 
to gauge boson pairs (only to quarks and charged leptons).
It does not modify the vacuum state considered, initially taken as having a spontaneously-broken supersymmetry
in the gauge-and-Higgs sector.
It breaks explicitly the $U(1)_A$ symmetry (\ref{uasym},\ref{ua0}), lifting the two pre\-viously-flat directions associated with $s_A$   and $A$.
With 
\be
\label{ma2}
\!\ba{ccl}
|\varphi_{\rm in}|^2 \!\!&=&\! |\, h_1 \sin\beta -\,h_2^c \cos\beta\,|^2 
\vspace{1mm}\\
\!&=&\!  
 |H^+|^2\! + \frac{1}{2}\,A^2 + 
 \frac{1}{2} \,|\sqrt 2  \,\hbox{\rm Re} \,( h_1^0 \,s_\beta -\,h_2^0 \,c_\beta)\,|^2,
\ea
\ee

\noindent
it provides an extra contribution $m_A^2$ to $m^2_{H^\pm}$,  
so that
\be
m^2_{H^\pm}= m_W^2 + m_A^2\,.
\ee

\vspace{.5mm}

Adding the supersymmetric $m_Z^2$ contribution associated with the $z$ in (\ref{expz}) 
and super\-sym\-me\-try-breaking contribution $m_A^2$ from   (\ref{ma0}) we get the scalar mass$^2$ matrix
\be
{\cal M}_\circ^2\,=\,\left(\ba{cc}
c_\beta^2\,m_Z^2+ s_\beta^2\,m_A^2 &-\,s_\beta c_\beta \, (m_Z^2+m_A^2)   \vspace{2mm}\\
- \,s_\beta c_\beta\,  (m_Z^2+m_A^2)   & s_\beta^2\,m_Z^2+ c_\beta^2\,m_A^2 
\ea\right)\,,
\ee
verifying
\be
\ba{ccc}
\hbox{Tr}\,{\cal M}_\circ^2\!&=&\! m_H^2+m_h^2 = m_Z^2 + m_A^2\,, 
\vspace{1mm}\\
\hbox{det}\,{\cal M}_\circ^2\!&=&\! m_H^2 \,m_h^2 = m_Z^2 \,m_A^2\,\cos^2\,2\beta\,,
\ea
\ee
\noindent
so that
\be
\ba{ccc}
m^2_{H,h}\!\!&=&\!\! \hbox{\small $\dis\frac{m_Z^2+m_A^2}{2}$}\pm 
\sqrt {\hbox{ \footnotesize$\dis\left(\frac{m_Z^2+m_A^2}{2}\right)^2 $}  -  m_Z^2 m_A^2\,\cos^22\beta }.
\vspace{-1mm}\\
\ea
\ee
It implies  $m_h < m_Z |\cos 2\beta |$ at the classical level, up to radiative corrections 
which must be significant  if one is to reach  $\simeq 125$ GeV/$c^2$
from a classical value below $m_Z$. 
These mass eigenstates behave for large $m_A$ as
\be
\left\{\ba{ccrcc}
H \!&\to&\! \sqrt 2 \ \,\hbox{\rm Re} \,( \, h_1^0 \,s_\beta -\,h_2^0 \,c_\beta)  \ \ \   (\hbox{large}\ m_H\simeq m_A)\,,
\vspace{1mm}\\
h&\to&\,h_{SM}\,=\, \sqrt 2 \ \,\hbox{\rm Re} \,( h_1^0 \,c_\beta +\,h_2^0 \,s_\beta)\ \ \  (\hbox{SM-like})\,.
\ea\right.
\ee

\vspace{1mm}
\noindent
The
$h$ field, presumably associated 
with the 125 GeV$/c^2$ boson observed at CERN, is then also very close to the $z$ in (\ref{expz})
for large $\tan\beta$, \,justifying {\it an almost  complete association of this 125 GeV$/c^2$ boson with the spin-1 $Z$}.

\vspace{-1mm}

\subsection{N/nMSSM}

\vspace{-2mm}

This also applies to extensions of the minimal model, as with an extra N/nMSSM singlet $S$
with a trilinear $\,\lambda\,H_1 H_2 S$ coupling,
 making it easier to get from $\lambda$  large enough spin-0 masses \footnote{$\lambda$ was denoted $h/\sqrt 2$ in \cite{R}, 
 \vspace{-.5mm}
 with 
$\lambda v/\sqrt 2 = hv/2\geq m_Z$ for $\lambda\geq \sqrt{(g^2+g'^2)/2}$. This allows for all spin-0 masses 
to be $\geq m_Z$ even before any breaking of the supersymmetry, {\it independently of $\tan \beta$}, 
in contrast with the MSSM.}.
In the N/nMSSM, first considered without a $\mu$ term, the supersymmetric contributions to spin-0 masses are \cite{R} 
\be
\left\{
\ba{l}
\ \ m_w = m_W, \ \ \ m_z= m_Z,\vspace{2mm}\\
m \ \hbox{\LARGE$ \left(\right.$}
\!\ba{c}
\vspace{-6mm}\\
\hbox{\small scalar}\ s_A,\ \hbox{\small pseudoscalar
\ A}
\vspace{0mm}\\
\hbox{\small complex singlet}
\ea \!
\hbox{\LARGE$ \left.\right)$}
\, = m_A= \hbox{\small$\dis \frac{\lambda  v }{\sqrt 2}$}\ .
\ea \right.
\ee
They correspond, already in the absence of supersymmetry breaking, to the neutral scalar doublet mass$^2$ matrix
\be
{\cal M}_\circ^2\,=\,\left(\ba{cc}
c_\beta^2\,m_Z^2+ s_\beta^2\,m_A^2 &s_\beta c_\beta \, (m_A^2-m_Z^2)  
\vspace{2mm}\\
s_\beta c_\beta \, (m_A^2-m_Z^2)   & s_\beta^2\,m_Z^2+ c_\beta^2\,m_A^2 
\ea\right)\,,
\ee
where $m_A={\lambda v}/{\sqrt2}\,$.

\vspace{1.5mm}

Adding as in (\ref{ma0}) the supersymmetry-breaking term
$-\,\delta m_A^2 \, | h_1 \,s_\beta -\,h_2^c \,c_\beta|^2$ $= -\,\delta m_A^2 \  |\varphi_{\rm in}|^2$
does not modify the vacuum state, while
shifting the $A$ and $w^\pm$ mass$^2$ by the same amount $\delta m_A^2$, so that
\be
\label{nmssmwa}
m_A^2= \hbox{\small$\dis \frac{\lambda^2 v^2 }{2}$} + \delta m_A^2,\ \ 
m_w^2 = m_W^2+\delta m_A^2 = m_W^2+ m_A^2- \hbox{\small$\dis \frac{\lambda ^2 v^2}{2}$}.
\ee
It provides as in the MSSM an extra contribution to the neutral scalar doublet mass$^2$ matrix, 
shifted by

\vspace{-4mm}

\be
\delta {\cal M}_\circ^2\,=\,\left(\!\ba{cc}
\ s_\beta^2\,\delta m_A^2 & \!\!-s_\beta c_\beta \,  \delta m_A^2
\vspace{1mm}\\
-s_\beta c_\beta \, \delta m_A^2   &  \,c_\beta^2\,\delta m_A^2
\ea\!\right)\,.
\ee
From this shift $m_A^2\! = \frac{\lambda^2 v^2 }{2} \to \frac{\lambda^2 v^2 }{2}+\delta m_A^2\,$,
the mass$^2$ matrix for the scalar doublet components reads
\be
\label{nmssm}
{\cal M}_\circ^2=\hbox{\small$\dis \left(\!\ba{cc}
c_\beta^2\,m_Z^2\!+ s_\beta^2\,m_A^2 & \!\!\!s_\beta c_\beta \, (\lambda^2 v^2-m_A^2\!-m_Z^2 )  
\vspace{2mm}\\
s_\beta c_\beta \, (\lambda^2 v^2-m_A^2\!-m_Z^2)   & \!\!\!s_\beta^2\,m_Z^2\!+ c_\beta^2\,m_A^2 
\ea\!\right)
$}.
\ee
For $\,\lambda\to 0\,$ $S$ decouples and the  spectrum
(\ref{nmssmwa},\ref{nmssm}) 
returns to the usual MSSM one. For $\,\lambda\neq 0\,$ further contributions involving also
a possible singlet mass term $\frac{1}{2}\,\mu_S\,S^2$
lead in general to a mixing between neutral doublet and singlet components, with
${\cal M}_\circ^2$ embedded into a $3\times 3$  matrix.

\vspace{-1mm}

\section{Conclusions}

\vspace{-2mm}

Independently of specific realisations
(MSSM, N/nMSSM, USSM, ...)
supersymmetric theories 
provide \hbox{spin-0} bosons as extra states for massive spin-1 gauge bosons,
despite different symmetry properties
and different couplings to quarks and leptons \cite{R,gh}. This further applies to supersymmetric grand-unified theories 
with extra dimensions \cite{gut,gutbis}.
By connecting spin-1 {\it mediators of gauge interactions} with \hbox{spin-0} {\it particles} associated with {\it symmetry breaking} 
and {\it mass generation},
supersymmetry provides an intimate connection between the electroweak gauge couplings  and the spin-0 couplings associated with symmetry breaking and mass generation.

\vspace{1.5mm}

The 125 GeV$/c^2$ boson recently observed at CERN may also be interpreted, up to a mixing angle induced by super\-symmetry breaking, as
the spin-0 partner of the $Z$ under {\it two} supersymmetry transformations,
\be
\label{gh3}
\hbox {\framebox [7.6cm]{\rule[-.18cm]{0cm}{.65cm} $ \dis
\hbox{\em spin-1} \ \,Z\ \  \ \stackrel{SUSY}{\longleftrightarrow }\ \ \stackrel{SUSY}{\longleftrightarrow }\  \ \hbox{\em spin-0 \,BEH boson}\,,
$}}
\ee
i.e.~as a $Z$ deprived of its spin.

\vspace*{3mm}

This provides within a theory of electroweak and strong interactions 

\vspace{-5mm}

\be
\ba{c}
\hbox{\it the first example of two known fundamental particles }
\vspace{1mm}\\
\hbox{\it 
of different spins related by supersymmetry},
\ea
\ee
 in spite of different electroweak properties. This is a considerable progress 
 as compared to the initial situation in
(\ref{hyp}), bringing further confidence in the relevance of supersymmetry 
for the description of fundamental particles and interactions.

\vspace{3mm}

{\it Supersymmetry may thus be tested in the gauge-and-BE-Higgs sector\,} at present and future colliders, in particular through the properties of the new boson, 
even if $R$-odd superpartners were 
still to remain out of reach for some time.

\vfill

\end{document}